\documentclass[twocolumn,showpacs,preprintnumbers,aps,amsmath,amssymb]{revtex4}

\usepackage{graphicx}
\usepackage{dcolumn}
\usepackage{bm}
\usepackage{epsfig}

\begin{document}


\title{Gr\"uneisen model for melts}

\author{U. Buchenau}
 \email{buchenau-juelich@t-online.de}
\affiliation{%
J\"ulich Center for Neutron Science, Forschungszentrum J\"ulich\\
Postfach 1913, D--52425 J\"ulich, Federal Republic of Germany
}%

\date{March 14, 2017}

\begin{abstract}
The Gr\"uneisen relation is shown to be important for the thermodynamics of dense liquids.  
\end{abstract}

\maketitle

One of the fascinating phenomena in dense liquids is the rapid disappearance of the liquid excess entropy (excess over the crystal) with decreasing temperature \cite{angellnew,angell}, leading to the definition of a fictive Kauzmann temperature below the kinetic freezing glass temperature.

On the experimental side, this disappearance has been observed in many systems \cite{angellnew}. It turns out to be well described in terms of the pragmatical Angell equation for the excess entropy of the liquid per particle \cite{angell}
\begin{equation}\label{kauz}
	S_1=S_{\infty,A}\left(1-\frac{T_K}{T}\right),
\end{equation}
where $T_K$ is the Kauzmann temperature at which the excess entropy becomes zero.

On the numerical side, eight years ago, Berthier and Tarjus \cite{tarjus,tarjus2} made the surprising discovery that the dynamics of the Lennard-Jones potential in the Kauzmann region changes drastically if one replaces its attractive tail by an external pressure (the WCA-potential \cite{wca}), though the pair correlation function remains practically unchanged.

The result motivated other groups to investigate the dependence of the dynamics on the specific form of the potential. Toxvaerd and Dyre \cite{toxvaerd} came to the conclusion that the dynamics of the dense liquid depends exclusively on the forces between nearest neighbors.

The Princeton group \cite{kauzmann} investigated modified Lennard-Jones potentials, in which the exponent of the repulsive part was varied from 12 down to 7, keeping the well depth constant. Among other results, they reported the temperature dependence of the purely configurational part of the entropy.

One can make a connection between the experimental and the numerical world by fitting these numerical results in terms of the pragmatical Angell expression, eq. (\ref{kauz}).

\begin{figure}[b]
\hspace{-0cm} \vspace{0cm} \epsfig{file=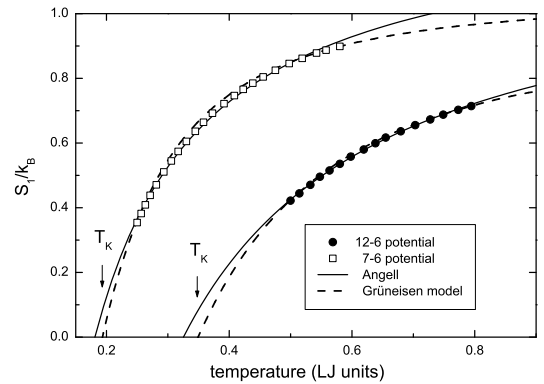,width=7 cm,angle=0} \vspace{0cm} \caption{Numerically determined structural single-particle entropies \cite{kauzmann} in the 12-6 and 7-6 Lennard-Jones potentials (full and open symbols, respectively). The continuous lines  are fits with the Angell \cite{angell} expression, eq. (\ref{kauz}), the dashed lines fits in terms of the Gr\"uneisen model of the present paper.}
\end{figure}

Fig. 1 shows the calculated \cite{kauzmann} structural entropies $S_1$ per particle in the 7-6 and in the 12-6 Lennard-Jones potential. The continuous lines are fits in terms of the Angell expression \cite{angell}
with two parameters, ($S_{\infty,A}=1.22k_B$, $T_K=0.32$ for the 12-6 and $S_{\infty,A}=1.33k_B$, $T_K=0.179$ for the 7-6). The dashed lines are fits in terms of the Gr\"uneisen model of the present paper.

One gets much closer to the Kauzmann temperature (to 1.38 $T_K$) in the 7-6 potential than in the 12-6 potential (1.54 $T_K$). This is strange, because it is decidedly the more harmonic potential of the two.

Setting the energy $\epsilon=1$ and the length $\sigma=1$, the 12-6 Lennard-Jones potential is given by
\begin{equation}
	V(x)=4\left(\frac{1}{x^{12}}-\frac{1}{x^6}\right)
\end{equation}
with the minimum at $r_0=2^{1/6}$. The 7-6 potential
\begin{equation}
	V(x)=4\left(\frac{2^{1/6}3}{x^7}-\frac{7}{2x^6}\right),
\end{equation}
has the same minimum and the same well depth. 

The Angell expression, eq. (\ref{kauz}), is notoriously difficult to explain. The equation tells us that one has to reckon with a finite structural entropy $S_{\infty,A}$. However, if one tries to describe the curves in Fig. 1 in terms of a gaussian in structural energy containing $\exp(S_{\infty,A}/k_B)$ states per atom, one fails; it is possible to get the right Kauzmann temperature by adapting the width of the gaussian, but the curvature that one gets is too small to reproduce the measured data.

If one thinks about it, one realizes that structural states with different energy should not only differ in the energy, but also in the equilibrium volume per atom and in the vibrational entropy. 

According to Toxvaerd and Dyre \cite{toxvaerd}, the Gr\"uneisen parameter which allows to determine these two quantities from the structural energy should be the one of the nearest neighbor potential.

In the melt, the three degrees of freedom per atom transform into three structural degrees of freedom. Each degree of freedom has a structural energy $E$, a vibrational entropy $S_E$ and a volume $V_E$ of the order of one third of the atomic volume, $V_0=V_a/3$.

One can obtain a relation between these three quantities via the Gr\"uneisen relation \cite{kittel} for the thermal volume expansion $\alpha$ of a solid 
\begin{equation}\label{gruen}
	\alpha=\frac{\gamma c_V}{BV_a}
\end{equation}
($\gamma$ Gr\"uneisen parameter, $c_V$ heat capacity per atom at constant volume, $B$ bulk modulus, $V_a$ atomic volume).

To translate this relation to the volume change $V_E$ on increasing the structural energy of a degree of freedom by $E$, one must remember that only half of the energy $c_VT$ goes into the potential; the other half is kinetic energy. This is different for the structural energy, which goes fully into the potential. Therefore
\begin{equation}
	\frac{V_E}{V_0}=1+\frac{2\gamma E}{BV_0}.
\end{equation}

According to the definition of the Gr\"uneisen $\gamma$, the volume change $V_E$ changes the Debye frequency by the factor $(V_0/V_E)^\gamma$ and the bulk modulus by the factor $(V_0/V_E)^{2\gamma}$. This implies an increase of the vibrational entropy by $\gamma k_B\ln(V_E/V_0)$ per degree of freedom. 

For what follows, one needs the behavior of the structure with a structural energy $E$ per degree of freedom under an applied pressure $p$ at the temperature $T$. It is practical to use the definitions $v_E=(V_E/V_0)$, $v_{E0}=1+2\gamma E/BV_0$ and $B_E=B(V_0/V_E)^{2\gamma}$. 

The volume at $p$ and $T$ is given by the minimum of the Gibbs free energy
\begin{equation}
	G(v_E)=\frac{1}{2}B_EV_0(v_E-v_{E0})^2-\gamma k_BT\ln{v_E}+pV_0v_E.
\end{equation}

This leads to a quadratic equation with the solution
\begin{equation}\label{ve}
	v_E=\frac{v_{E0}-p/B_E}{2}+\sqrt{\frac{(v_{E0}-p/B_E)^2}{4}+\frac{\gamma k_BT}{B_EV_0}}.
\end{equation}

For simplicity, let us describe the $Z_0=\exp(S_\infty/k_B)$ structural states per degree of freedom in the liquid in terms of a constant density of states $Z_0/E_M$ between the structural energy zero and the maximum energy $E_M$.

It is instructive to imagine what will happen in a liquid at constant volume. The state at $E_M$ will tend to gain vibrational entropy by expanding. This will lead to an entropic pressure which compresses the state at zero energy to a lower volume. Thus one gets different volumes for different states even at constant volume.

The results \cite{kauzmann} in Fig. 1 were obtained at constant volume, enforced by an external pressure and by the attractive forces of second and higher neighbors. 

However, if one follows Toxvaerd and Dyre \cite{toxvaerd} and considers only the forces between nearest neighbors, one can describe it by a pressure $p$ which acts only on the first coordination shell and is zero at temperature zero.

For a given potential, one can calculate the Gr\"uneisen $\gamma$ from the ratio $V''/V'''$ of the second and third derivatives of the potential \cite{kittel} at the atomic distance $r_0$
\begin{equation}\label{gv}
	\gamma=\frac{r_0V'''(r_0)}{6V''(r_0)},
\end{equation}
yielding $\gamma=2.66$ for the 7-6 potential and 2.95 for the 12-6 potential. With twelve neighbors (six bonds per atom, two bonds per degree of freedom), one calculates the values for $BV_0$ in Table I.

The structural part of the liquid entropy is calculated from the partition function
\begin{equation}
	Z_l=\frac{Z_0}{E_M}\int_0^{E_M}\exp(-E/k_BT+\gamma\ln{v_E})dE.
\end{equation}

\begin{table}[htbp]
	\centering
		\begin{tabular}{|c|c|c|c|c|c|}
\hline
substance             & $\gamma$   &$S_\infty/k_B$&  $BV_0$  & $E_M$    & $T_K$     \\
\hline   
12-6 Lennard-Jones    & 2.95       &   0.31       &  12.7    & 0.89     & 0.355     \\
7-6 Lennard-Jones     & 2.66       &   0.348      &  7.4     & 0.6      & 0.195     \\
selenium              & 1.37       &   0.25       &  11.3    & 1.3      & 0.355     \\
\hline		
		\end{tabular}
	\caption{Melt parameters of the Gr\"uneisen model in Lennard-Jones units.}
	\label{tab:Comp}
\end{table}

At constant volume, it is first necessary to calculate the temperature-dependent pressure which keeps the system at constant volume. Doing this for the two Lennard-Jones potentials, one finds the plausible result that the vibrational entropy averages to zero. This implies that the entropy difference between liquid and crystal at constant volume is purely configurational.

The resulting curves in Fig. 1, obtained with the parameters $S_\infty$ and $E_M$ in Table I, describe the numerical data satisfactorily, supporting the Gr\"uneisen interpretation.

The ratio of the two $E_M$- and $T_K$-values in Table I is close to the ratio of the curvatures of the two potentials at $r_0$, in accordance with the empirical Lindemann criterion of melting \cite{lindemann}. $3S_\infty$ is close to Richards rule, according to which the melting entropy per atom of a simple liquid is 1.1 $k_B$, and also close to the melting entropy of 0.8 to 0.9 $k_B$ at constant volume \cite{ulf}. The Angell values $S_{\infty,A}$ turn out to be a factor of 1.3 larger than the $3S_\infty$-values of the Gr\"uneisen model.

In real experiments, the pressure is practically zero and one has thermal expansion. One can calculate an effective Gr\"uneisen parameter of the structural degrees of freedom from 
\begin{equation} \label{dcda}
	\gamma=\frac{\Delta\alpha BV_a}{2\Delta c_p},
\end{equation}
where $\Delta\alpha$ and $\Delta c_p$ are the differences in the thermal volume expansion coefficient and the heat capacity per atom, respectively, measured at the glass transition between liquid and glass.

\begin{figure}[b]
\hspace{-0cm} \vspace{0cm} \epsfig{file=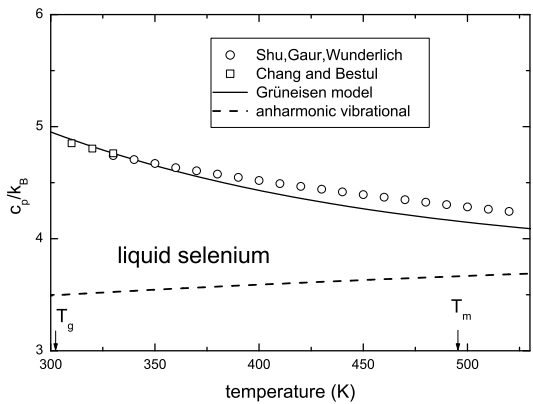,width=7 cm,angle=0} \vspace{0cm} \caption{Measured \cite{chang,gaur} excess heat capacity data in liquid selenium. The continuous line is calculated with the Gr\"uneisen model, taking all parameters from other experiments. The dotted line shows the anharmonic vibrational part.}
\end{figure}

Fig. 2 shows a fit of heat capacity data \cite{chang,gaur} of liquid selenium with the parameters in Table I. 

The value of $\gamma$ in Table I was obtained from eq. (\ref{dcda}). With the values of $\Delta c_p$, $\Delta\alpha$ and $BV_a$ measured at the glass transition \cite{gaur,simha}, one finds $\gamma=1.37$, a bit higher than the Gr\"uneisen $\gamma=1$ found from the thermal expansion of the glass \cite{grosse}.

In order to translate the real selenium into the Lennard-Jones world, the Kauzmann temperature $T_K=240$ K, determined by Angell \cite{angell2} from crystal and liquid heat capacity data \cite{chang}, was set equal to the Kauzmann temperature of the 12-6 potential in Table I. $S_\infty$ was taken to be three quarters of the average value of the two Lennard-Jones potentials, because selenium has only nine nearest neighbors, two covalent and seven van-der-Waals ones \cite{knudsen}.

With these parameters fixed, the product $BV_0$ was chosen in such a way that its average value calculated in the Gr\"uneisen model at the pressure zero corresponded (in units of $k_BT_K$) to the one measured \cite{simha} at the glass temperature in the glass phase.

Two small contributions from other sources \cite{gaur} were added to the calculated value, namely the anharmonic vibrational component 0.12$k_BT/T_g$ from the glass expansion and a contribution of 0.24$k_B$ from the polymer-ring conversion, which at these comparatively low temperatures is a constant.  

The total anharmonic vibrational part obtained in this way is a bit larger than the one determined independently from the softening of neutron spectra \cite{se}, but still agrees within experimental error.

Selenium demonstrates that one needs to do the calculation per degree of freedom and not per atom. If one calculates per atom, one does not get good agreement. But this is also intuitively evident: A structural degree of freedom moves many atoms.

Eq. (\ref{dcda}), which relates $\Delta\alpha$ to $\Delta c_p$, can be inserted into the Prigogine-Defay relation \cite{jackle,gundermann} 
\begin{equation}\label{prigo}
	\frac{\Delta c_p\Delta\kappa}{(\Delta\alpha)^2V_aT_g}=\Pi,
\end{equation}
with $\kappa$ compressibility. $\Pi$ is always larger than 1 for glass transitions. This yields
\begin{equation}\label{prig}
	\frac{\Delta\kappa}{\kappa}=4\gamma^2\Pi\frac{\Delta c_pT}{BV_a}.
\end{equation}

J\"ackle's interpretation \cite{jackle} of the Prigogine-Defay ratio predicts $\Pi=1$ if all volume changes at pressure zero have the same volume-energy ratio, i.e. the same $\gamma$. For a collection of different $\gamma$-values, J\"ackle's equs. (2a), (2b) and (2c) yield in a straightforward way
\begin{equation}\label{pri}
	\Pi=\frac{\overline{\gamma^2}}{\overline{\gamma}^2},
\end{equation}
with $\overline{\gamma}$ as the proper $\gamma$ for the calculation of the thermal expansion.

Selenium has a Prigogine-Defay ratio of two \cite{simha}. This is consistent with the neutron results \cite{se}, which show a strong anharmonicity for the van-der-Waals springs and practically no anharmonicity for the two covalent springs per atom. The van-der-Waals $\gamma$ must be close to three, in order to get the average one up to the value 1 seen in the glass \cite{grosse}.

Obviously, some of the structures have a higher and some have a lower van-der-Waals content in their energy, giving them a higher or a lower $\gamma$ than the average one. The structural degrees of freedom cannot be identical with the vibrational ones; otherwise their average $\gamma$ should be one and the Prigogine-Defay ratio should be three.

In the Lennard-Jones liquids, one stands a much better chance to find a Prigogine-Defay ratio of one and, consequently, isomorphism \cite{ulf}. 

If one fits the data in Fig. 2 with the Angell expression, eq. (\ref{kauz}), one finds $S_{\infty,A}=2.4k_B$. This value is now a factor of three larger than $3S_\infty$, about a factor of two more than in the Lennard-Jones fits, probably because the temperature development is faster at constant pressure and because there is an appreciable amount of vibrational entropy. 

Molecular glasses \cite{angell}, where eq. (\ref{kauz}) was first proposed, have even larger values of $S_{\infty,A}$, a factor of three to seven higher than the one in selenium.

Naturally, molecules like salol and orthoterphenyl have much more degrees of freedom than a single atom. But it is still surprising to see an average factor of five. It implies that not only the molecular orientation, but also inner degrees of freedom participate in the creation of structural entropy. One can obviously build more structures by deforming the molecule itself. 

The example of selenium shows that the structural energy goes also into the hard degrees of freedom, though to a lesser extent. This is consistent with the factor of three to seven in molecular glasses, which is always smaller than the number of atoms per molecule. 

Finally, let us return to the surprising difference between the 12-6 and the WCA potential \cite{tarjus,tarjus2}.

As shown above, the Kauzmann temperature $T_K$ is roughly proportional to the curvature of the potential at the minimum. One can use this proportionality to calculate an effective and temperature-dependent Kauzmann temperature for the WCA-potential (a 12-6 Lennard-Jones potential with a removed attractive part \cite{wca}). 

Of course, to keep nearest neighbors at $r_0$ for any nonzero temperature with the WCA potential, one has to apply a pressure. At the density 1.1, this pressure extrapolates approximately to zero at the temperature zero \cite{tarjus2}. It is at this density where the bifurcation of the dynamics of the two potentials is strongest.

The strongly different dynamics was first found \cite{tarjus} at a temperature of 0.45 Lennard-Jones units. With six nearest-neighbor-bonds per atom, one has approximately a potential energy of 1/4 $k_BT$ per bond. In the WCA-potential, one then needs a force of 3 Lennard-Jones units to keep the atom at $r_0$. With this force, the calculated mean square displacement at this temperature is a factor of 3/2 larger than the one in the 12-6 Lennard-Jones without force, meaning that the dynamics corresponds to the one of the 12-6 potential at a factor of 3/2 higher temperature.

At lower temperatures, the effective Kauzmann temperature of the WCA sample with density 1.1 gets lower and lower, while the one of the 12-6 potential stays constant. Consequently, the difference in the dynamics gets larger and larger \cite{tarjus2}.

To conclude, it was shown that one needs the Gr\"uneisen concept to understand the thermodynamics of dense liquids. 

The basis for this work was laid by Andrew Phillips, who realized that one could calculate the anharmonic vibrational entropy from the neutron spectra of selenium, twenty eight years ago.

\end{document}